\documentclass[a4paper]{jpconf}
\usepackage{hyperref}
\usepackage{graphicx}
\begin{document}
\title{ArDM: a ton-scale liquid Argon experiment for direct detection
of Dark Matter in the Universe}

\author{Andr\'e Rubbia}

\address{Institut f\"{u}r Teilchenphysik, ETHZ, CH-8093 Z\"{u}rich, Switzerland}

\ead{andre.rubbia@cern.ch}

\begin{abstract}
The ArDM project aims at developing and operating large noble liquid
detectors to search for direct evidence of Weakly Interacting 
Massive Particle (WIMP) as Dark Matter in the Universe. 
The initial goal is to design, assemble and operate
a $\approx$1~ton liquid Argon prototype to demonstrate the feasibility
of a ton-scale experiment with the required performance to
efficiently detect and sufficiently discriminate backgrounds 
for a successful WIMP detection. Our design addresses the
possibility to detect independently ionization and scintillation
signals.
In this paper, we describe this goal and the conceptual design
of the detector.
\end{abstract}

\section{Introduction}
Astronomical observations give strong evidence for the existence of
non-luminous and non-baryonic matter, presumably composed of a new
type of elementary particle. A leading candidate is the
Weakly Interacting Massive Particle (WIMP). If they exist, they should
form a cold thermal relic gas which can be detected via elastic collisions
with nuclei of ordinary matter.
The detection of these WIMPs is based on the capability of measuring the recoils of target nuclei with kinetic energy in the range of
$10-100$~keV. 
The signal is therefore quite elusive and it is a rare event given the weak coupling.
Furthermore the rate is not easily predicted, 
since it depends on many poorly defined variables, even in the context of well defined extensions of the SM 
like e.g. SUSY~\cite{kaufmann}. Nonetheless, ton-scale targets are nowadays to be 
contemplated in order study with high statistical power the DAMA result\cite{Bernabei:2004yy} or alternatively
to cover large fractions of the remaining theoretical parameter space.

Within the ICARUS R\&D program,  it was first shown that noble liquid 
detectors using Xenon or Argon could act as targets for WIMP detection~\cite{ica1,Arneodo:2000vc}.
Xenon or Argon provide a high event rate because of their high density 
and high atomic number and large target masses are readily conceivable. 
They have high scintillation and ionization yields because of 
their low ionization potentials. Both scintillation and ionization 
are measurable and can be used to very effectively discriminate 
between nuclear recoils and gamma/electron backgrounds.

\def\R2Lurl#1#2{\mbox{\href{#1}{\tt #2}}}               

The use of noble liquid gases to detect WIMP dark matter is currently the subject
of intense R\&D carried out by a number of groups worldwide\cite{Aprile:2005mz,Spooner:2001br,Brunetti:2004rk}.
In these detectors, one relies on the simultaneous detection of the ionization
charge and of the scintillation light produced during a nuclear recoil event. 
A main subject for any such detector is the method of the readout for the ionization
and scintillation. Currently, the XENON\cite{Aprile:2005mz}, ZEPLIN\cite{Spooner:2001br} 
and WARP\cite{Brunetti:2004rk} designs rely exclusively on
photomultipliers (PMTs) for their readout. 
The possibility to directly detect the ionization charge is less well developed although it might provide alternative
and potentially large benefits. Given the low energy thresholds necessary to efficiently
detect WIMP signals, this method however requires the charge to be amplified
before it is readout. While amplification is not possible in the liquid Argon phase, it
can be achieved in the vapor in equilibrium on top of the liquid, 
although operation in this context precludes
the inclusion of common avalanche quenchers, since they will condense in
the liquid phase. 

In 2004 we have initiated the Argon Dark Matter experiment 
(ArDM\footnote{ETH Z\"urich, Granada University,
CIEMAT, Soltan Institute Warszawa, Z\"urich University.}, 
see \R2Lurl{http://neutrino.ethz.ch/ArDM/ }{http://neutrino.ethz.ch/ArDM/}). 
The goal of this project is to design, assemble and operate
a bi-phase $\approx$1~ton Argon detector with independent
ionization and scintillation readout, to demonstrate the feasibility
of a noble gas ton-scale experiment with the required performance to
efficiently detect and sufficiently discriminate backgrounds 
for a successful WIMP detection. 

\section{Conceptual design}
\begin{figure}[htb]
\begin{minipage}{18pc}
\includegraphics[width=18pc]{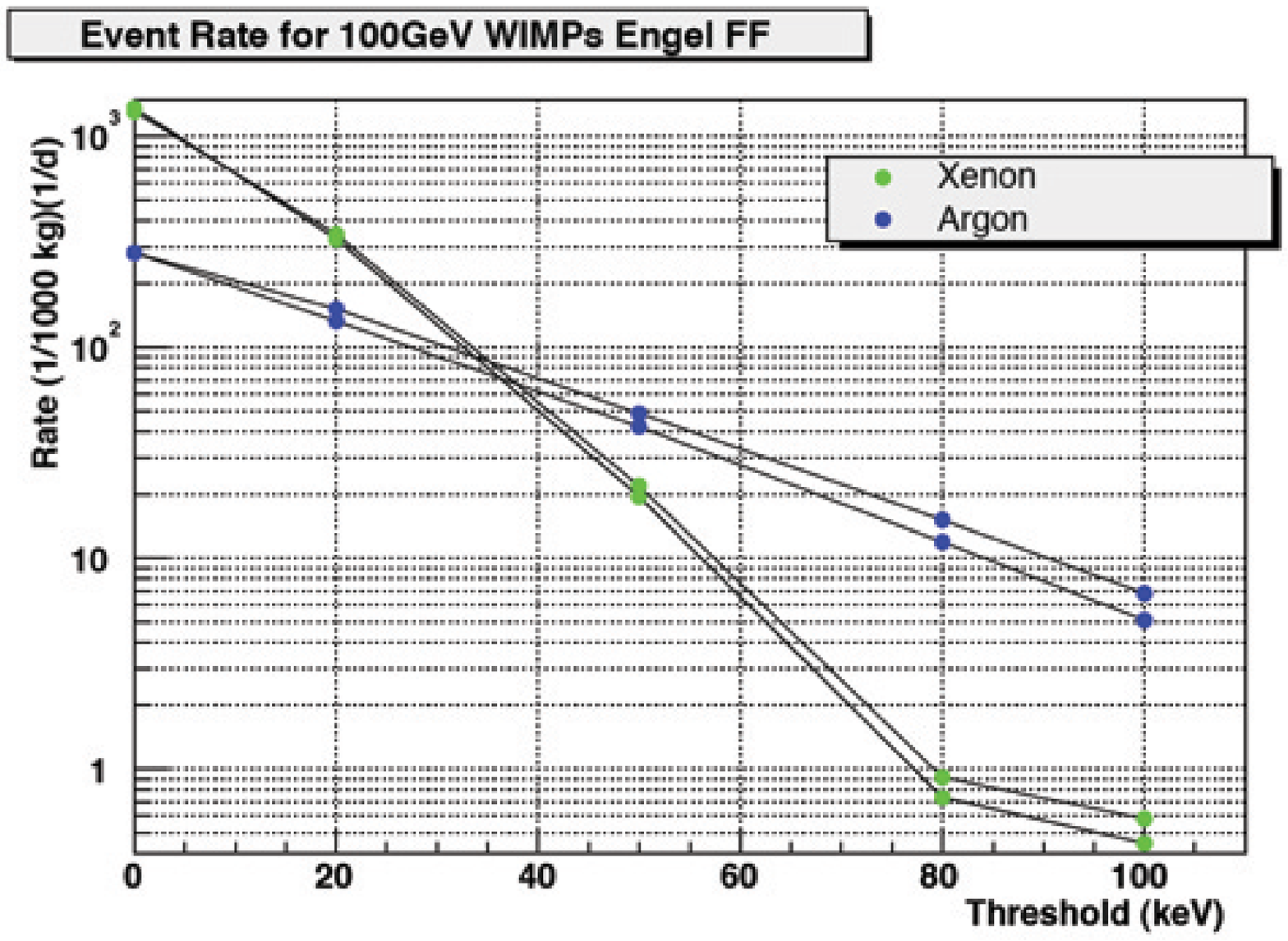}
\caption{\label{fig:recoil}Event rate per ton per day on Argon and Xenon targets 
for winter and summer periods as a function of detection threshold 
for a WIMP-nucleon cross-section of 10$^{-42}$ cm$^{2}$ = 10$^{-6}$ pb 
and a WIMP mass of 100 GeV.}
\end{minipage}\hspace{2pc}%
\begin{minipage}{18pc}
\includegraphics[width=18pc]{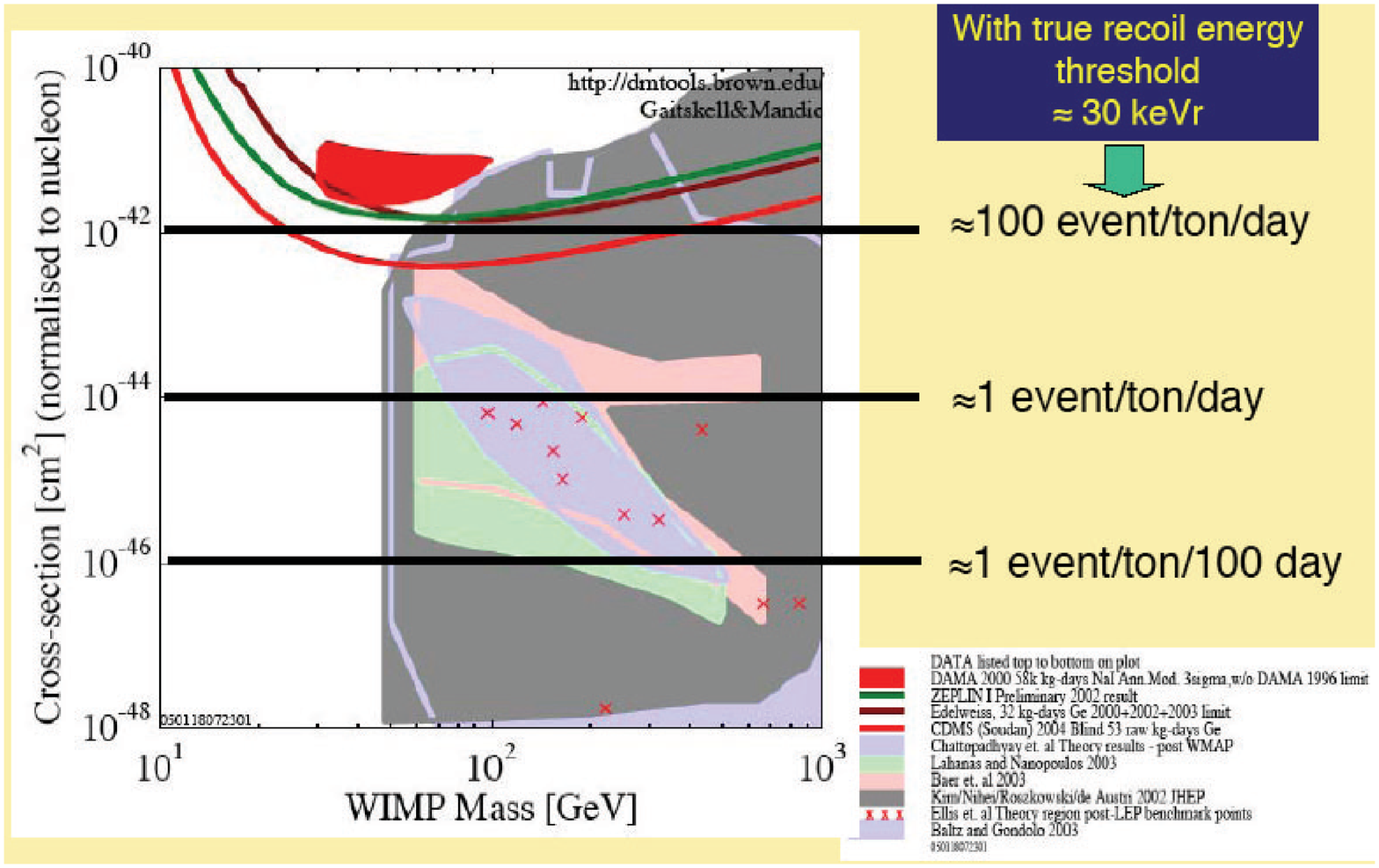}
\caption{\label{fig:eventrate}Cross-section normalized to nucleon
versus WIMP mass. The expected event rates for a true recoil energy threshold 
of 30 keVr are indicated by horizontal thick lines. 
With such a threshold a WIMP-nucleon cross-section 
of 10$^{-44}$ cm$^{2}$ yields one event per ton per day.}
\end{minipage} 
\end{figure}

The choice of natural Argon for the initial ton-scale target instead of Xenon 
can be motivated by three arguments:\\
(1) The detection energy threshold depends on the achievable performance of the light
and ionization detection systems. The event rate in Argon is less sensitive to the threshold 
on the recoil energy than for Xenon because of form factors. 
For a threshold of \ensuremath{\approx}30 keVr, the rates on 
Xenon and Argon per mass are similar (See 
Figure~\ref{fig:recoil}). With such a threshold a WIMP-nucleon cross-section 
of 10$^{-44}$ cm$^{2}$ yields about one event per ton per day (See 
Figure~\ref{fig:eventrate}).\\
(2) Argon is much cheaper than other noble gases, and we have 
acquired sizeable experience in the handling of massive liquid 
Argon detectors within the ICARUS program. A ton-scale Argon detector
is hence readily conceivable, safe and economically affordable.\\
(3) The scientific relevance of obtaining data on Argon and Xenon 
is given by the fact that recoil spectra in Xenon and Argon are 
different (due to kinematics), providing an important crosscheck 
in case of a positive signal.

One non-negligible drawback of natural Argon liquefied from the 
atmosphere is the existence of the radioactive isotope $^{39}$Ar 
which is a beta-emitter with a lifetime of 269 years and a value 
Q=565{\nobreakspace}keV. Its concentration in atmospheric Argon is well 
known~\cite{loosli} and will induce a background decay rate of \ensuremath{\approx}1 kHz 
in a 1{\nobreakspace}ton detector. In principle, the intrinsic electron/nuclear 
recoil rejection provided by the ratio of the scintillation to 
the ionization yields, which is extremely high for nuclear recoils 
(i.e. WIMP events), is sufficient to suppress this background, provided
this ratio can be measured precisely.
This fact needs to be experimentally further understood since rejection
factors exceeding $>10^9$ are needed. We 
intend to fully address it with our proposed 1 ton prototype, i.e. a detector
of the relevant size\footnote{We note that achieving the required performance
on small prototypes is less challenging.}. 
We are also studying other ways to obtain $^{39}$Ar-depleted targets, 
by using Argon extracted from well gases (extracted from underground 
natural gas) rather than from the atmosphere. This would provide 
a reduction of this background although its cost is to be estimated. 
On the other hand, the $^{39}$Ar decays, evenly distributed in the 
target, provide a precise calibration and monitoring of the detector 
response as a function of time and position.

\begin{figure}
\begin{center}
\includegraphics[width=0.95\textwidth]{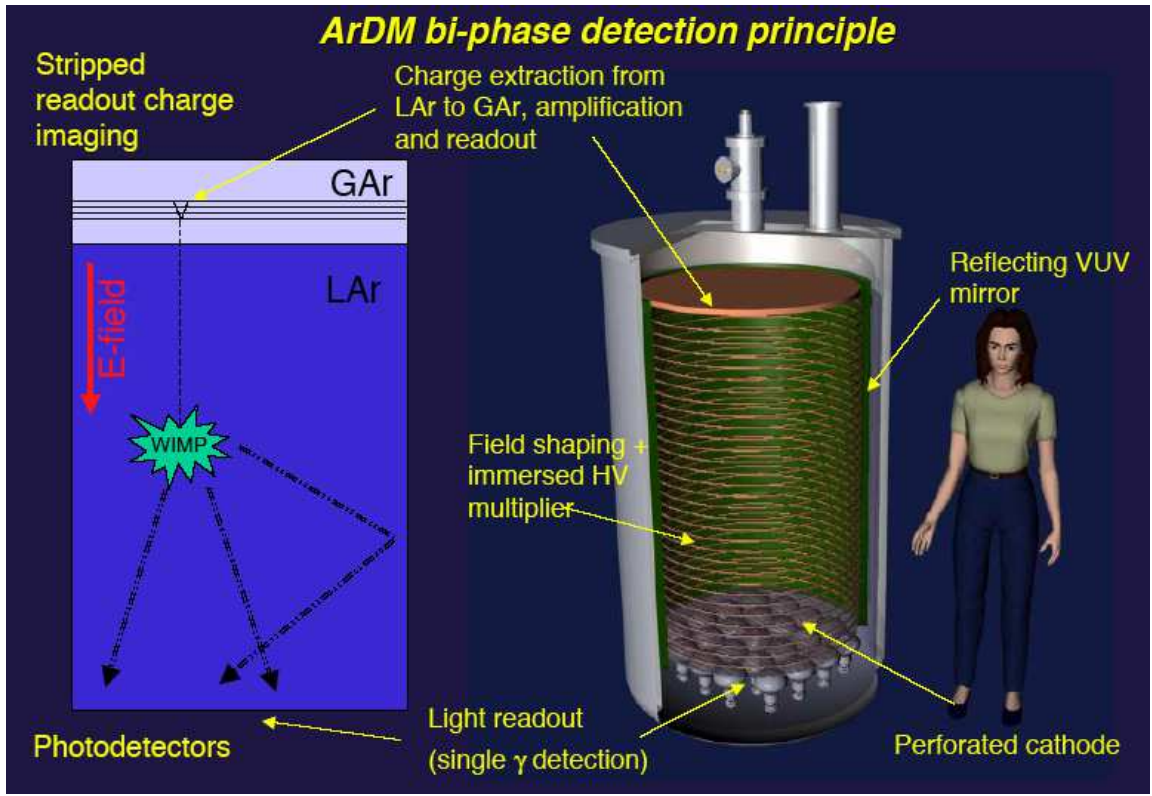}
\end{center}
\caption{\label{fig:concept}The conceptual design of the ArDM experiment.}
\end{figure}

The conceptual layout of the detector is shown in Figure~\ref{fig:concept}. 
More details can be found in Ref.~\cite{como}.
A main feature is the possibility to independently detect the ionization 
charge and scintillation light. 
Following an ionizing event, ionization charges will be drifted 
towards the top of the detector where they will be extracted 
from the liquid to the gas phase. There, a Large Electron Multiplier\cite{LEM,Chechik:2004wq} 
(LEM) system will amplify the electrons in order to produce a 
detectable signal. By segmenting the LEMs, an image of the event 
will be obtained, retaining the salient features of the ICARUS 
imaging technology, although with a much lower energy threshold.

Because background discrimination requires the ratio of the scintillation 
to the ionization yields, the primary VUV scintillation light 
of argon (128 nm) will be reflected by specially conceived high 
reflectivity mirrors made
of $Al-MgF_2$ coated Mylar foils\cite{knecht} and
located on the field shaping electrodes. The photons are
detected via a light readout system located 
behind the transparent HV cathode. R\&D efforts are under way to 
improve on the light collection efficiency (about 5\% with PMTs), 
and hence on the threshold and background discrimination, by 
using wavelength shifters and alternative light readout systems 
such as avalanche photodiodes.

Charge imaging and time correlation between scintillation and 
charge will provide a precise localization of the event vertex 
(in space), hence a good fiducial volume definition, important 
for \ensuremath{\gamma}-ray and slow neutrons background rejection from 
surrounding elements. 

The time dependence of scintillation light can be used to further 
discriminate between heavy recoils and other backgrounds (in 
addition to primary versus secondary signal).

A second feature of the experiment is the possibility to reach very high drift 
fields up to 5~kV/cm in order to detect an ionization signal 
even in the presence of highly quenched nuclear recoils as in 
the case of a WIMP interaction.

\section{Outlook}
A natural follow-up of the use of liquid noble gases as media 
for detectors is the extension of their application to the direct 
detection of nuclear recoils induced by dark matter. We have 
presented our plans for the construction of a 1 ton prototype 
whose goal is to demonstrate the validity of our design. This 
goal requires a successful implementation and operation of
(a) a high drift field device;
(b) a LEM based charge readout;
(c) a highly efficient Argon scintillation light detection system.

Given the challenging nature of the experiment 
which requires innovations both at the level of the detection 
methods and at the level of background rejection, our immediate 
plan is to fully design and acquire the needed equipments to 
setup and operate the 1 ton prototype at CERN. The operation 
of the prototype will involve cryogenic, LAr purification, HV 
system, drift volume, charge amplification + readout, and light 
readout. It will allow us to define and set up all the necessary 
equipment and infrastructure for a safe operation of the detector.

Our first milestone is a proof of principle and stability studies, 
and further optimization of the design for a highly efficient \ensuremath{\gamma}-ray 
and beta electron ($^{39}$Ar) rejection vs. nuclear recoils. Strong 
neutron shielding and stringent requirements on detector radio-purity 
will be fully addressed in a second phase.
Assuming the successful operation of the prototype, we will consider 
a deep underground operation\footnote{A memorandum has been sent on September 9th, 2005
to the Canfranc Scientific Committee.}.
With the assumed recoil energy threshold of 30 keVr, a WIMP-nucleon 
cross-section of 10$^{-42}$~cm$^{2}$ would yield 100 events per day 
per ton (See Figure ~\ref{fig:eventrate}). The sensitivity expectation of the ArDM 
1 ton prototype would therefore be around \ensuremath{\sim}10$^{-6}$ pb 
or better. The discovery region of the ArDM 1 ton detector, assuming 
that sufficiently low gamma and neutron backgrounds can be reached, 
would be \ensuremath{\sim}10$^{-8}$ pb. Its ultimate sensitivity for a year 
of operation would be \ensuremath{\sim}10$^{-10}$ pb. Scaling linearly with 
mass, a \ensuremath{\approx}10 ton detector would reach \texttt{<}10$^{-11}$~pb 
in a year of operation.

\section*{Acknowledgments}
The help of all ArDM colleagues from ETH Z\"urich, Granada University,
CIEMAT, Soltan Institute Warszawa, and Z\"urich University, is greatly acknowledged.
Informal contributions from P.~Picchi (LNF) are also greatly recognized.

\smallskip

\end{document}